\documentclass[pra,showpacs,twocolumn,superscriptaddress]{revtex4-1}
\usepackage{bm,graphicx,amsmath}

\newcommand{\beq}{\begin{equation}}
\newcommand{\enq}{\end{equation}}

\begin{document}
\title{Chaos-driven dynamics in spin-orbit coupled atomic gases}
\author{Jonas Larson}
\email{jolarson@fysik.su.se}
\affiliation{Department of Physics, Stockholm University, Se-106 91 Stockholm, Sweden}
\affiliation{Institut f\"ur Theoretische Physik, Universit\"at zu K\"oln, K\"oln, De-50937, Germany}
\author{Brandon M. Anderson}
\affiliation{Joint Quantum Institute, national Institute of Standards and technology and the University of Maryland, Gaithersburg, Maryland 20899-8410, USA}
\author{Alexander Altland}
\affiliation{Institut f\"ur Theoretische Physik, Universit\"at zu K\"oln, K\"oln, De-50937, Germany}
\date{\today}

\begin{abstract}
The dynamics, appearing after a quantum quench, of a trapped, spin-orbit coupled, dilute atomic gas is studied. The characteristics of the evolution is greatly influenced by the symmetries of the system, and we especially compare evolution for an isotropic Rashba coupling and for an anisotropic spin-orbit coupling. As we make the spin-orbit coupling anisotropic, we break the rotational symmetry and the underlying classical model becomes chaotic; the quantum dynamics is affected accordingly. Within experimentally relevant time-scales and parameters, the system thermalizes in a quantum sense. The corresponding equilibration time is found to agree with the Ehrenfest time, i.e. we numerically verify a $\sim\log(\hbar^{-1})$ scaling. Upon thermalization, we find the equilibrated distributions show examples of quantum scars distinguished by accumulation of atomic density for certain energies. At shorter time-scales we discuss non-adiabatic effects deriving from the spin-orbit coupled induced Dirac point. In the vicinity of the Dirac point, spin fluctuations are large and, even at short times, a semi-classical analysis fails.
\end{abstract}


\maketitle

\section{Introduction}
The physics of ultracold atomic gases has greatly advanced in recent years~\cite{zwerger}. The high control of system parameters, together with the isolation of the system from its environment, have made it possible to use such setups to simulate various theoretical models of condensed matter physics~\cite{zwerger,bloch}. Of significance in many condensed matter models is the response to external magnetic fields. Since atoms are neutral, there is no direct way to implement a Lorentz force in these systems. Early experiments created a synthetic magnetic field via rotation~\cite{pethick}. While simple theoretically, these methods are impractical for certain setups, and they are limited to weak, uniform fields. The first experimental demonstration of {\it laser-induced synthetic magnetic fields} for neutral atoms~\cite{spielman1}, on the other hand, paves the way for an avenue of new situations to be studied in a versatile manner~\cite{ohberg1,FluxLattice,PlanarBField}. Owing to numerous fundamental applications in the condensed matter community~\cite{SORev,TIRev}, maybe the most important direction appears when the laser fields induce a synthetic spin-orbit (SO) coupling. Indeed, a certain kind of SO-coupling for neutral atoms has already been demonstrated~\cite{spielman2}, and it is expected that more general SO-couplings will be attainable within the very near future~\cite{Npod,Spherical}.

While SO-couplings can in principle bear identical forms in condensed matter and cold atom models, there is an inevitable difference, often overlooked, between these two systems. The presence of a confining potential for the atomic gas can qualitatively change the physics~\cite{zwerger,pethick}, and has only recently been addressed~\cite{trap1,trap2,TrappedSOC1,TrappedSOC2, shenoy}. Furthermore, most of these studies are concerned with ground/stationary state properties of the system~\cite{trap1,TrappedSOC1,TrappedSOC2}, while few works discuss dynamics or non-equilibrium physics. Notwithstanding, the experimental isolation of these systems suggests that they are well suited for studies of {\it closed quantum dynamics}~\cite{polkovnikov}.  

Historically, some of the finest experiments regarding dynamics of closed quantum systems have been performed in quantum optics~\cite{haroche,cqed}. An early example proved quantization of the electromagnetic field by making explicit use of quantum revivals~\cite{walther}. Such quantum recurrences, in general connected to integrability or small system sizes, are now well understood. The situation becomes more complex for non-integrable systems~\cite{polkovnikov} or systems with a large number of degrees-of-freedom~\cite{drummond}. One particularly interesting question is whether any initial state relaxes to an asymptotic state, and if so, what are then the properties of this ``equilibrated'' state and the mechanism behind the equilibration. Both these questions have inspired numerous publications during the last decade, both theoretical~\cite{thermo1,thermo2} as well as experimental~\cite{thermoexp,thermoexp2,thermoexp3}. A rule of thumb is that in order for a closed quantum system to {\it thermalize}, i.e. all expectation values can be obtained from a microcanonical state, its underlying classical Hamiltonian should be non-integrable~\cite{polkovnikov}. While true in most cases studied so far, exceptions to this hypothesis has been found~\cite{eisert}. Moreover, the behavior near the transition from regular to chaotic dynamics, classically explained by Kolmogorov-Arnold-Moser theory~\cite{kamtheory}, is not well understood for a quantum system~\cite{kam}. It is therefore desirable to study a system where these two regimes can be explored by tuning an external parameter, and for which the experimental methods in terms of preparation and detection are already well developed. 

Motivated by the above arguments, in this paper we consider dynamics of a trapped SO-coupled cold dilute atomic gas. The SO-coupling is assumed tunable from isotropic (Rashba-like) to anisotropic, and hence the system can be tuned between regular and chaotic. Note that even though this crossover is generated by a change in the form of the SO-coupling, the confining trap causes the system to become non-integrable. We distinguish between short and long time evolution, where by ``long time'' we mean times similar to the Ehrenfest time. In fact, the corresponding time-scale for the thermalization is found to agree with the Ehrenfest time, and thereby scale as $\log(\hbar^{-1})/\lambda$ where $\lambda$ is the maximum Lyaponov exponent. This scaling for the thermalization has been conjectured in Ref.~\cite{altland}, but was not numerically verified in these works. At shorter times when the wave packet remains localized, we especially study the rapid changes in the spin as the wave packet evolves in the vicinity of the Dirac point (DP).  For energies below the DP ($E<0$), we utilize an adiabatic model derived in the Born-Oppenheimer approximation (BOA)~\cite{boa}.  Aside from some special initial states, we encounter thermalization in all cases. These exceptions correspond to states evolving within a regular ``island'' in the otherwise chaotic sea. Among the thermalized states, the equilibrated distributions are found to show {\it quantum scars} originating from periodic orbits of the underlying classical model. The experimental relevance of all our theoretical predictions are discussed and put in a state-of-the-art experimental perspective. 

The paper is outlined as follows. The following section introduces the system Hamiltonian and discusses its symmetries. Section~\ref{ssec2b} derives the adiabatic model by imposing the BOA. A semi-classical analysis, demonstrating classical chaos for anisotropic SO-couplings, is presented in Sec.~\ref{sec3}. The following section considers the full quantum model at short times, Sec.~\ref{ssec4a}, and long times, Sec.~\ref{ssec4b}.  Section~\ref{ssec4c} contains a discussion regarding experimental relevance of our results. Finally, Sec.~\ref{sec5} gives some concluding remarks.

\section{Spin-orbit coupled cold atoms}\label{sec2}

\subsection{Model spin-orbit Hamiltonian}\label{ssec2a}

Several proposals exist for implementing spin-orbit couplings in cold atoms~\cite{Nlevel,NonAbelian,dynamic_so}. In general, these synthetic spin-orbit fields are generated through the application of optical and Zeeman fields to produce a set of dressed states that are well separated energetically from the remaining dressed states~\cite{ohberg1}. We denote these states as pseudo-spin, but emphasize that there is no connection to real space rotations. Spatial variation of the dressed states will couple the pseudo-spin to the orbital motion of the atom. An atom prepared in a pseudo-spin state will therefore see an effective Hamiltonian, provided the atom is sufficiently cold. 

For a specific configuration of optical fields, one can induce the effective Hamiltonian~\cite{NonAbelian}
\begin{equation}\label{SOham}
\hat{H}_{SO} =\frac{\hat{\mathbf{p}}^2}{2m} + \frac{1}{2} m \omega^2 \mathbf{r}^2 +v_x\hat{p}_x\hat{\sigma}_x+v_y\hat{p}_y\hat{\sigma}_y,
\end{equation}
where $\hat{\mathbf{p}} = (\hat{p}_x, \hat{p}_y)$ is the momentum operator, $\hat{\mathbf{r}} = (\hat{x}, \hat{y})$ is the position operator, $m$ is the mass of the atom, and $\omega$ the frequency of a harmonic trap. The operator $\hat{\sigma}_i$ is the $i$-th Pauli matrix in pseudo-spin space, and the velocities $v_i$ couple pseudo-spin to an effective momentum dependent Zeeman field, $\mathbf{B}(\mathbf{p}) = (v_x p_x, v_y p_y)$. This momentum-dependent Zeeman field can simulate any combination of the Rashba \cite{rashba} and Dresselhaus \cite{dresselhaus} SO-couplings experienced in semiconductor quantum wells and systems alike. 

In the absence of a trap, $\omega=0$, the spectrum of (\ref{SOham}) is
\begin{equation}
E_\mu(p_x, p_y) = \frac{1}{2m}\left( p_x^2 + p_y^2 \right) + \mu \sqrt{(v_x p_x)^2 + (v_y p_y)^2} 
\end{equation}
with the corresponding eigenfunctions
\begin{equation}
\left|{\psi_{\mu, \mathbf{p}}}\right> = e^{i m (v_x x + v_y y)} 
\left|\varphi_\mu\right>
\end{equation}
where 
\begin{equation}\label{adstate}
\left|\varphi_\mu\right>
=\frac{1}{\sqrt{2}}\left(
e^{-i\varphi/2}|\uparrow\rangle - \mu e^{i\varphi/2}|\downarrow\rangle\right),
\end{equation}
is a spinor with helicity $\mu=\pm1$ and $\varphi=\arctan(v_yp_y/v_xp_x)$. These states have well defined momentum, but have no velocity since $\left< \dot{\mathbf{r}} \right> = \left< \nabla_\mathbf{p} H \right> = 0$, provided the optical fields are maintained. Note further that the eigenstates are parametrically dependent on $p_x$ and $p_y$.

We remark that for an isotropic SO-coupling, $v_x=v_y$, the Hamiltonian (\ref{SOham}) is equivalent to the dual $E\times\varepsilon$ Jahn-Teller model, frequently appearing in chemical/molecular physics and condensed matter theories~\cite{Exe}. With a simple unitary rotation of the Pauli matrices, the SO-coupling attains the more familiar Rashba form~\cite{rashba} (or equivalently Dresselhaus form~\cite{dresselhaus}). For $v_x\neq v_y$, i.e. when the SO-coupling is anisotropic, the model becomes the dual $E\times(\beta_x+\beta_y)$ Jahn-Teller model~\cite{Exe}. In particular, the $\hat{z}$-projection of total angular momentum, $\hat{J}_z=\hat{L}_z+\frac{\hat{\sigma}_z}{2}$, is a constant-of-motion for the isotropic but not for the anisotropic model. More precisely, breaking of the SO isotropy implies a reduction in symmetry from $U(1)$ to $Z_2$.

Throughout we will use dimensionless parameters where the oscillator energy $E_o=\hbar\omega$ sets the energy-scale, $l=\sqrt{\hbar/m\omega}$ the length-scale, and the characteristic time is $\tau=\omega^{-1}$. We note that for typical experimental setups, $\omega\sim 10-100\,\, \textrm{Hz}$ and $m (v_x^2 +v_y^2) / \hbar \sim 1-10\,\, {\rm kHz}$. Moreover, in what follows we will refer to pseudo-spin simply as spin.  When necessary, we introduce a parameter ${h}$ serving as a dimensionless Planck's constant, i.e. $h\hbar$. In this way, $h$ controls the strength of Planck's constant and by varying it we can explore how the dynamics depends on $\hbar$. 

\subsection{Adiabatic model}\label{ssec2b}

The large ratio of the SO energy to trapping energy, typically $mv^2 / \hbar \omega \sim 10 - 1000$, suggests that a BOA~\cite{boa} will be valid for experimental implementations. The separation of timescales of the spin and orbital degrees of freedom implies that in some regimes we can factorize the wavefunction as the product of spin and orbital wavefunctions. A spin initially aligned with the adiabatic momentum-dependent magnetic field ${\bf B}({\bf p})$ will remain locked to that field at future times, provided the center of mass motion avoids the DP. We then solve for the spin wavefunction at an instantaneous orbital configuration and use this answer to find an adiabatic potential for the orbital motion. This is in analogy with the traditional BOA, where the electronic and nuclear wavefunctions are approximated as a product, and the electron degrees of freedom instantaneously adjust to the adiabatic potential given by the nuclear degrees of freedom.

\begin{figure}
\includegraphics[width=5cm]{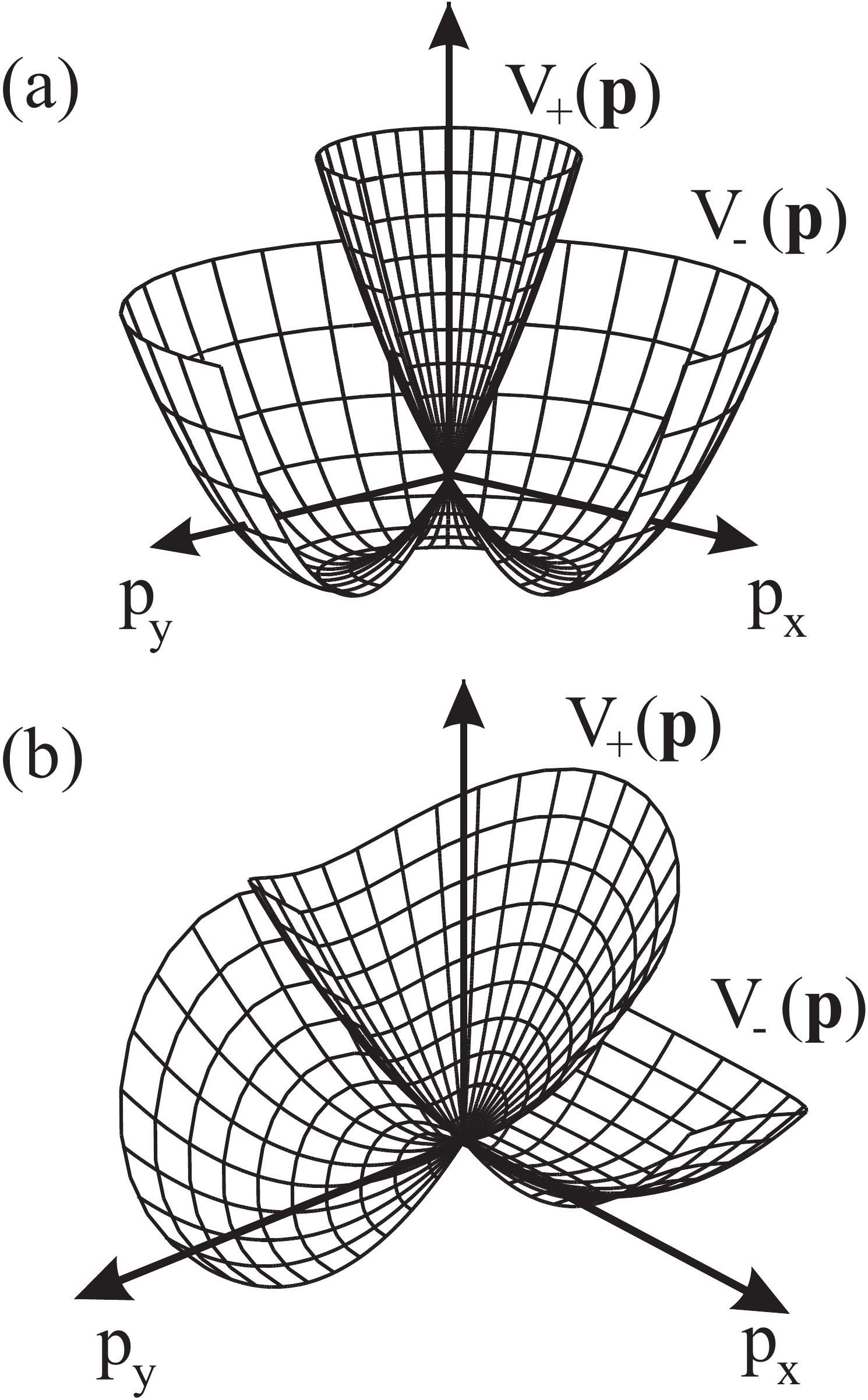}
\caption{Adiabatic potentials of the isotropic (a) and anisotropic (b) SO-coupled models. In both figures, the $E=0$ plane is the one including the DP at $p_x=p_y=0$. A necessary, but not sufficient, condition for the validity of the BOA is that $E<0$. In (a), the lower adiabatic potential $V_-(p_x,p_y)$ has the characteristic sombrero shape. By considering an anisotropic SO-coupling, the rotational symmetry is broken and $V_-(p_x,p_y)$ possesses two global minima at $(p_x,p_y)=(0,\pm v_y)$. }
\label{fig1}
\end{figure}

In our BOA, we have chosen the {\it adiabatic states}~\cite{boa} for the orbital motion to be the spin-helicity states, given by (\ref{adstate}). If we project the Hamiltonian into the basis $|\varphi_\mu\rangle$, we arrive at the adiabatic potential
\begin{equation}\label{adham}
\hat{H}_{\rm ad}^{(\mu)} =\frac{\hat{x}^2}{2}+\frac{\hat{y}^2}{2}+\frac{\hat{p}_x^2}{2}+\frac{\hat{p}_y^2}{2}+\mu\sqrt{v_x^2\hat{p}_x^2+v_y^2\hat{p}_y^2}.
\end{equation}
The trap thus takes the role of kinetic energy and (\ref{adham}) can be pictured as a particle in a (dual) {\it adiabatic potential}
\begin{equation}
V_\mu(\hat{p}_x,{\hat{p}_y)=\frac{\hat{p}_x^2}{2}+\frac{\hat{p}_y^2}{2}+\mu\sqrt{v_x^2\hat{p}_x^2+v_y^2\hat{p}_y^2}}.
\end{equation}
shown in Fig.~\ref{fig1} for both the isotropic (a) and anisotropic (b) cases. 
We have neglected non-adiabatic corrections arising from the vector potential and the Born-Huang term~\cite{bornhuang}. For example, an additional scalar potential 
\begin{equation}\label{nonadcoup}
V_{\rm nad}(p_x,p_y) \sim \frac{(v_x v_y)^2 (p_x^2+p_y^2)}{\left(v_x^2p_x^2+v_y^2p_y^2\right)^2}.
\end{equation}
will emerge from the action of the SO-coupling on the spinor $\left|\varphi_\mu\right>$. This term is order $V_{\rm nad} \sim \left< \varphi \right| \nabla_{\bf p}^2 \left| \varphi \right> \sim 1/p^2$. There will also be an additional vector potential term $A \sim 1/p$. The non-adiabatic corrections diverge near the DP, but then fall off rapidly at finite $p$.  The adiabatic approximation, i.e. BOA, will be valid if the particle avoids $p=0$. We will show later that this condition is met if the particle is in the lower band, $\mu=-1$, and has energy $E<0$. 

Imposing the BOA, any state propagating on the lower adiabatic potential will be denoted $\Phi(p_x,p_y,t)$, and it is understood that 
\begin{equation}\label{adstate2}
\Phi(p_x,p_y,t)=\phi(p_x,p_y,t)|\varphi_-\rangle.
\end{equation}
The real space wave function $\Psi(x,y,t)$ is given as usual from the Fourier transform of $\phi(p_x,p_y,t)$. The time-evolution follows from $\phi(p_x,p_y,t)=\exp\left(-i\hat{H}_{ad}^{(-)}t\right)\phi(p_x,p_y,0)$. It is also clear that the state $\Phi(p_x,p_y,t)$ determines the spin orientation which is inherent in the ket-vector $|\varphi_i\rangle$. More explicitly, the time-evolved Bloch vector
\begin{equation}\label{blochvec}
\mathbf{R}(t)=\left(R_x(t),R_y(t),R_z(t)\right)\equiv\left(\langle\hat{\sigma}_x\rangle,\langle\hat{\sigma}_y\rangle,\langle\hat{\sigma}_z\rangle\right)
\end{equation}
takes the form
\begin{eqnarray}
R_x(t) & = & \int dp_xdp_y\,|\phi(p_x,p_y,t)|^2\cos(\varphi), \nonumber \\
R_y(t) & = &\int dp_xdp_y\,|\phi(p_x,p_y,t)|^2\sin(\varphi), \label{blochvec} \\
R_z(t) & = & 0 \nonumber
\end{eqnarray}
in the BOA, and it is remembered that the parameter $\varphi$ depends on $p_x$ and $p_y$. Note that the Bloch vector precesses in the equatorial spin $xy$-plane. If the wave packet $\Phi(p_x,p_y,t)$ is sharply localized, a crude approximation for the Bloch vector is given by 
\begin{eqnarray}\label{blochapp}
\bar{R}_x(t) & = &\frac{v_x\bar{p}_x(t)}{\sqrt{\left(v_x\bar{p}_x(t)\right)^2+\left(v_y\bar{p}_y(t)\right)^2}}, \\
\bar{R}_y(t) & = & \frac{v_y\bar{p}_y(t)}{\sqrt{\left(v_x\bar{p}_x(t)\right)^2+\left(v_y\bar{p}_y(t)\right)^2}}, \\
\bar{R}_z(t) & = & 0,
\end{eqnarray}
where $\bar{p}_\alpha(t)=\int dp_xdp_y\,|\Phi(p_x,p_y,t)|^2p_\alpha$ with $\alpha=x,\,y$. 

\section{Classical dynamics}\label{sec3}

Quantum chaos is often defined by having an underlying chaotic classical model. For the full model~(\ref{SOham}), the spin degrees-of-freedom cannot be eliminated in a straightforward manner in the vicinity of the Dirac point and as a consequence it is not {\it a priori} clear what the underlying classical model would be in this regime. On the other hand, in the BOA, the adiabatic Hamiltonian $\hat{H}_{ad}^{(-)}$ can serve as our classical model Hamiltonian. Still, it should be noted that we assume $\langle\hat{H}_{ad}^{(-)}\rangle\ll0$, such that the spectrum contains a sufficiently large number of energies below $E=0$. Furthermore, we point out that justification of the BOA does not necessarily imply approval of a semi-classical approximation which depends on the system energy and the actual shape of the dual potential $V_-(p_x,p_y)$. Nevertheless, as we will demonstrate in the following, for the chosen parameters, the agreement is indeed very good.

\begin{figure}
\includegraphics[width=8cm]{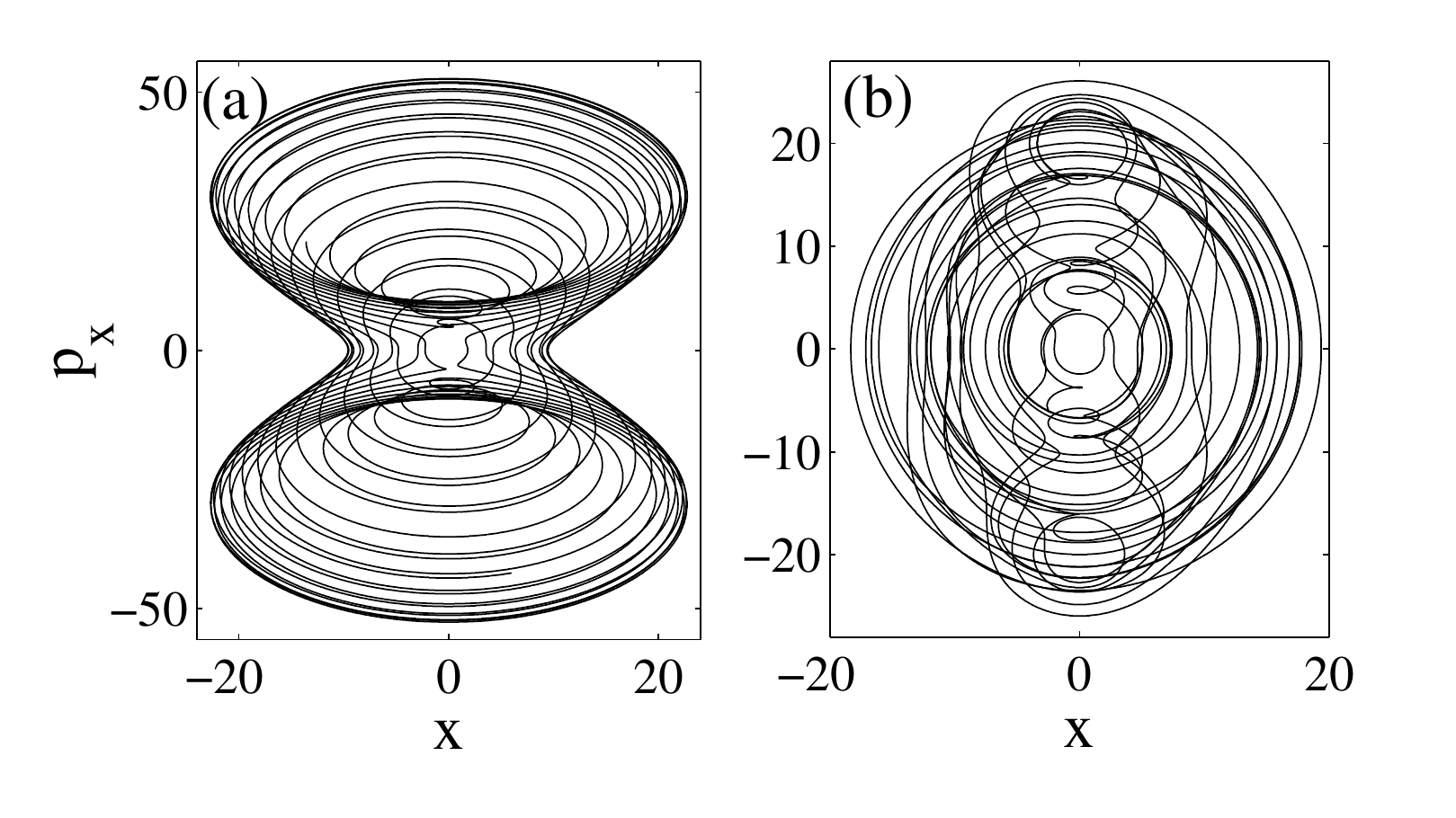}
\caption{Two examples of classical trajectories $((x(t),P_x(t))$ for regular (a) and chaotic (b) dynamics. In (a), typical for regular motion the trajectories evolve upon a tori. Contrary, in (b) the trajectory is much more irregular which is characteristic for the chaotic evolution. The regular motion is calculated for the SO-coupling strengths $v_x=v_y=30$, and the chaotic motion with $v_x=20$ and $v_y=30$. In both cases, the energy is $E=-192$.}
\label{fig_revision}
\end{figure}  

\begin{figure}
\includegraphics[width=8cm]{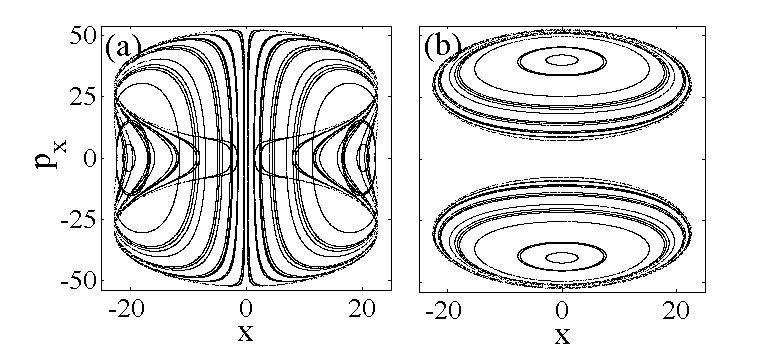}
\caption{Poincar\'e sections of the Rashba SO-coupled adiabatic model (\ref{adham}) for the intersections $y=0$ (a) and $p_y=0$ (b). The initial energy is $E=-192$, the SO-coupling strengths $v_x=v_y=30$, and the number of simulated semi-classical trajectories 18.}
\label{fig2}
\end{figure} 

The classical equations-of-motion of the Hamiltonian $\hat{H}_{ad}^{(-)}$ are
\begin{eqnarray}\label{heiseom}
\dot{x} & = &p_x-\frac{v_x^2p_x}{\sqrt{v_x^2p_x^2+v_y^2p_y^2}}, \\
\dot{p}_x &= &-x, \\
\dot{y} & = &p_y-\frac{v_y^2p_y}{\sqrt{v_x^2p_x^2+v_y^2p_y^2}}, \\
\dot{p}_y &= &-y.
\end{eqnarray}
For the Rashba SO-coupling, $v_x=v_y=v$, there is one unstable fix point $(p_x,p_y)=(0,0)$ and a seam of stable fix points $p_x^2+p_y^2=v^2$, see Fig.~\ref{fig1} (a). For the anisotropic case, $v_y>v_x$, there are three unstable fix points, $(p_x,p_y)=(0,0)$ and $(p_x,p_y)=(\pm v_x,0)$, while there are two stable fix points $(p_x,p_y)=(0,\pm v_y)$, see Fig.~\ref{fig1} (b).

The classical energy $E(x,p_x,y,p_y)=p_x^2/2+p_y^2/2+x^2/2+y^2/2-\sqrt{v_x^2p_x^2+v_y^2p_y^2}$ determines a hypersurface in phase space for any given energy $E(x,p_x,y,p_y)=E_0$. The semi-classical trajectories $(x(t),p_x(t),y(t),p_y(t))$ live on this surface. For the integrable case, $v_x=v_y$, these surfaces form different tori characteristic for quasi-periodic motion. As the rotational symmetry is slightly broken, $v_x\neq v_y$, the tori deforms and the motion loses its quasi-periodic structure~\cite{kamtheory}. This is the generic crossover from regular to chaotic classical dynamics. As an example of this generic behavior, we show in Fig.~\ref{fig_revision} two randomly sampled trajectories in the $xp_x$-plane for regular (a) and chaotic (b) evolution. For all results of this section, we solve the set of coupled differential equations~(\ref{heiseom}) using the Runge-Kutta (4,5) algorithm modified by {\it Gear's method}, suitable for stiff equations. We have also numerically verified our results employing different algorithms~\cite{alg}. As will be discussed further below, even in the chaotic regime, periodic orbits may persist and will greatly affect the dynamics, both at a classical and a quantum level~\cite{scar}. Such orbits are not, however, visible from Fig.~\ref{fig_revision}. 

\begin{figure}
\includegraphics[width=8cm]{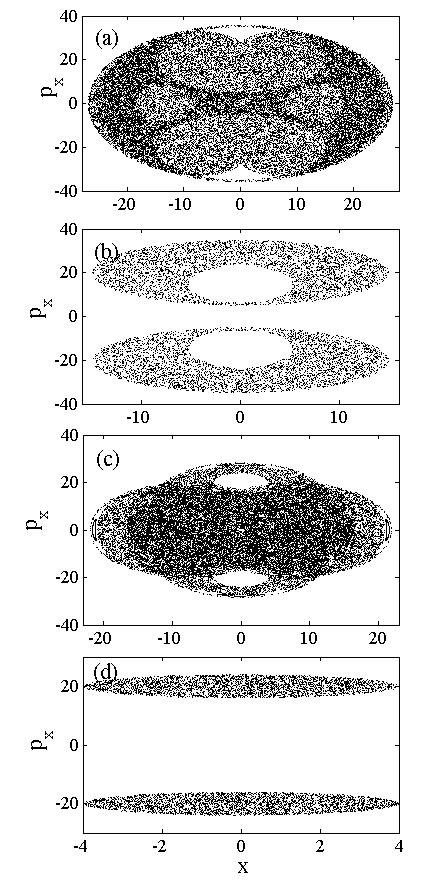}
\caption{Poincar\'e sections of the anisotropic SO-coupled adiabatic model (\ref{adham}) for $y=0$ (a) and (c), and for $p_y=0$ (b) and (d). The initial energies are $E=-88$ (a) and (b), and $E=-192$ (c) and (d), and the SO-coupling strengths $v_x=20$ and $v_y=30$ for both cases. The corresponding maximum Lyaponov exponents have been derived to $\lambda\approx0.12$ and $\lambda=0.090$ respectively. The number of semi-classical trajectories is the same as for Fig.~\ref{fig2}, namely 18.}
\label{fig3}
\end{figure}

The semi-classical behavior of classical dynamical systems is favorable visualized using Poincar\'e sections~\cite{strogatz}. Corresponding sections for the system (14)-(17) are depicted in Figs.~\ref{fig2} and \ref{fig3}. In the first figure we display the Poincar\'e sections in the $xp_x$ plane for the intersections determined by $y=0$ (a) or $p_y=0$ (b) of the isotropic model with the SO-coupling amplitudes $v_x=v_y=30$. The initial energy is taken as $E=-192$, well below the DP, consistent with the BOA. In (b), the section defined by $p_y=0$, the evolution results in ellipses in the Poincar\'e section, characteristic of quasi periodic motion. The structure of the Poincar\'e section for $y=0$ (a) is somewhat more complex. This can be understood from the sombrero shape of the adiabatic potential $V_-(p_x,p_y)$; for given $x=x'$, $p_x=p_x'$, $y=0$, and energy $E_0$, there are four possible values of $p_y$, and this multiplicity of possible $p_y$'s allow the ``curves'' in Fig.~\ref{fig2} (a) to cross. It should be noted that any single curve does not cross itself. Furthermore, by adding the $p_y$ values to Fig.~\ref{fig2} we have verified that neither of the corresponding three dimensional curves cross. 

Figure~\ref{fig3} presents two examples for anisotropic SO-couplings, both with $v_x=20$ and $v_y=30$. The quasi-periodic evolution is lost and the dynamics become mixed, with regions of both chaos and regular dynamics. The same conclusions were found in Ref.~\cite{jtchaos3} where a related Jahn-Teller model was studied. The two lower plots consider the same energies as in Fig.~\ref{fig2}, i.e. $E=-192$, while for (a) and (b) $E=-88$. Expectedly, the higher energy increases the accessible volume of phase space. For both energies we find {\it islands} free from chaotic trajectories. As will be demonstrated in the next section, within these islands the evolution is regular and the system does not thermalize. The plots also demonstrate clear structures also appearing in the chaotic regimes of the Poincar\'e sections in which the density of solutions changes. 


\section{Quantum dynamics}\label{sec4}
The idea of this section is to analyze how the corresponding quantum evolution is affected by whether the classical dynamics is regular or chaotic. Of particular importance is the long time evolution in which the system state may or may not equilibrate. However, we study also the short time dynamics arising for a localized wave packet traversing the Dirac point. In this regime, clearly the classical results of the previous section does not hold.

To study the system beyond the classical approximation, we solve the time-dependent Schr\"odinger equation, represented by the Hamiltonians~(\ref{SOham}) or (\ref{adham}), to obtain the corresponding wave function $\Psi(x,y,t)$ at time $t$. Note that for the full model~(\ref{SOham}), the wave function contains the spin degree-of-freedom $\Psi(x,y,t)=\psi_\uparrow(x,y,t)|\uparrow\rangle+\psi_\downarrow(x,y,t)|\downarrow\rangle$. 
The non-equilibrium initial state appears after a quench in the center of the trap. We prepare the system in a quasi-ground state for a shifted trap, and at $t=0$ suddenly move the trap center to $x_s=y_s=0$,
\begin{widetext}
\begin{equation}\label{shiftpot}
V(x,y)=\frac{(x-x_s)^2}{2}+\frac{(y-y_s)^2}{2},\hspace{0.7cm}\left\{\begin{array}{lll}x_s\neq0\,\,\mathrm{and/or}\,\,y_s\neq0, & & \mathrm{t}<0,\\ x_s=y_s=0, & & \mathrm{t}\geq0.\end{array}\right.
\end{equation}
\end{widetext}
By ``quasi-ground state'' in an anisotropic SO-coupled system, we consider an initial state predominantly populated in one of the two minima of the adiabatic potential $V_-(p_x,p_y)$. This seems experimentally reasonable where small fluctuations will favor one of the two minima. For the isotropic case, the phase of $\Phi(p_x,p_y,t=0)$ is taken randomly in agreement with symmetry breaking. Given the evolved states $\Psi(x,y,t)$, we are interested in the Bloch vector~(\ref{blochvec}) or its components, and the distributions $|\Phi(p_x,p_y,t)|^2$ and $|\Psi(x,y,t)|^2$. 

The numerical calculation is performed employing the {\it split-operator method}~\cite{split} which relies on factorizing, for short times $\delta t$, the time-evolution operator into a spatial and a momentum part. For small SO-couplings $v_x$ and $v_y$, the method is relatively fast, while as $v_x$ and/or $v_y$ are increased the time-steps $\delta t$ must be considerably reduced and the necessary computational power rises rapidly. In addition, for large $v_x$ and $v_y$, the grid sizes of position and momentum space must be increased, which also increases the computation time. Thus, we will limit the analysis to SO-couplings $v_x,\,v_y\leq30$. Furthermore, we have found by convergence tests that the full model (\ref{SOham}) requires much smaller time-steps $\delta t$ than the adiabatic one (\ref{adham}), and most of our simulations will therefore be restricted to energies $E<0$ for which the BOA is justified. 

The full quantum simulations are complemented by the semi-classical truncated Wigner approximation (TWA), which has turned out very efficient in order to reproduce quantum dynamics~\cite{twa}. The TWA considers a set of $N$ different initial values $(x_i,y_i,p_{xi},p_{yi})$ randomly drawn from the distributions $|\Psi(x,y,0)|^2$ and $|\Phi(p_x,p_y,0)|^2$. These are then propagated according to the classical equations-of-motion (\ref{heiseom}). The propagated set $(x_i(t),y_i(t),p_{xi}(t),p_{yi}(t))$ gives the semi-classical distributions, from which expectation values can be evaluated.

\subsection{Short time dynamics}\label{ssec4a}
Before investigating the prospects of thermalization, we first consider {\it short time} dynamics, by which we mean time-scales where the wave packet remains localized. In this respect, it is tempting to think of the dynamics as semi-classical. However, in the vicinity of the the DP any classical description would fail. Equivalently, the spin degrees-of-freedom will show large fluctuations which are difficult to capture classically. The short time dynamics is consequently most interesting for situations with energies $E>0$ where both the semi-classical approximation and the BOA break down, implying that the simulation is performed using the full model Hamiltonian~(\ref{SOham}). For these energies, the wave packet can traverse the DP and population transfer between the two adiabatic potentials $V_\mu(\hat{p}_x,\hat{p}_y)$ typically occurs. It is known that such non-adiabatic transitions can play important roles for the dynamics, and that the actual transition probabilities between the two potentials may be extremely sensitive to small fluctuations in the state~\cite{altland,dong}. In this subsection we especially address such non-adiabatic effects.

There are indeed several relevant time-scales in the dynamics: $(i)$ The spin precession time $T_{sp}$ gives the typical time for spin evolution and is proportional 
to the effective magnetic field $|{\bf B}({\bf p})|$, $(ii)$ the classical oscillation period $T_{cl}=2\pi$, and $(iii)$ the thermalization time $T_{th}$, which estimates the time it takes for the system to thermalize, i.e. when expectation values become approximately time independent. Normally, the magnitudes of these times follow the list above (in growing order), except in the vicinity of the DP where $T_{sp}\sim T_{cl}$ or even $T_{sp}\ll T_{cl}$ very close to the DP. While the first two are well defined, defining the last one is non-trivial. We can say that $(i)$ and $(ii)$ characterizes short time-time scales, and $(iii)$ long time-scales. As will be numerically demonstrated, the thermalization time turns out to scale as $\log(h^{-1})/\lambda$, where $h$ is the effective dimensionless Planck's constant and $\lambda$ the maximum Lyaponov exponent. This suggests that the thermalization time agrees with the Ehrenfest time 
\begin{equation}\label{ehrenfest}
T_E=\log(V/h)/\lambda,
\end{equation}
with $V$ the effective occupied phase space volume. $T_E$ is also the typical time-scale where semi-classical (TWA) expectation values no longer agree with quantum expectation values, which can be seen as a breakdown of Ehrenfest's theorem~\cite{duffing_classlim}. 

\begin{figure}
\includegraphics[width=8cm]{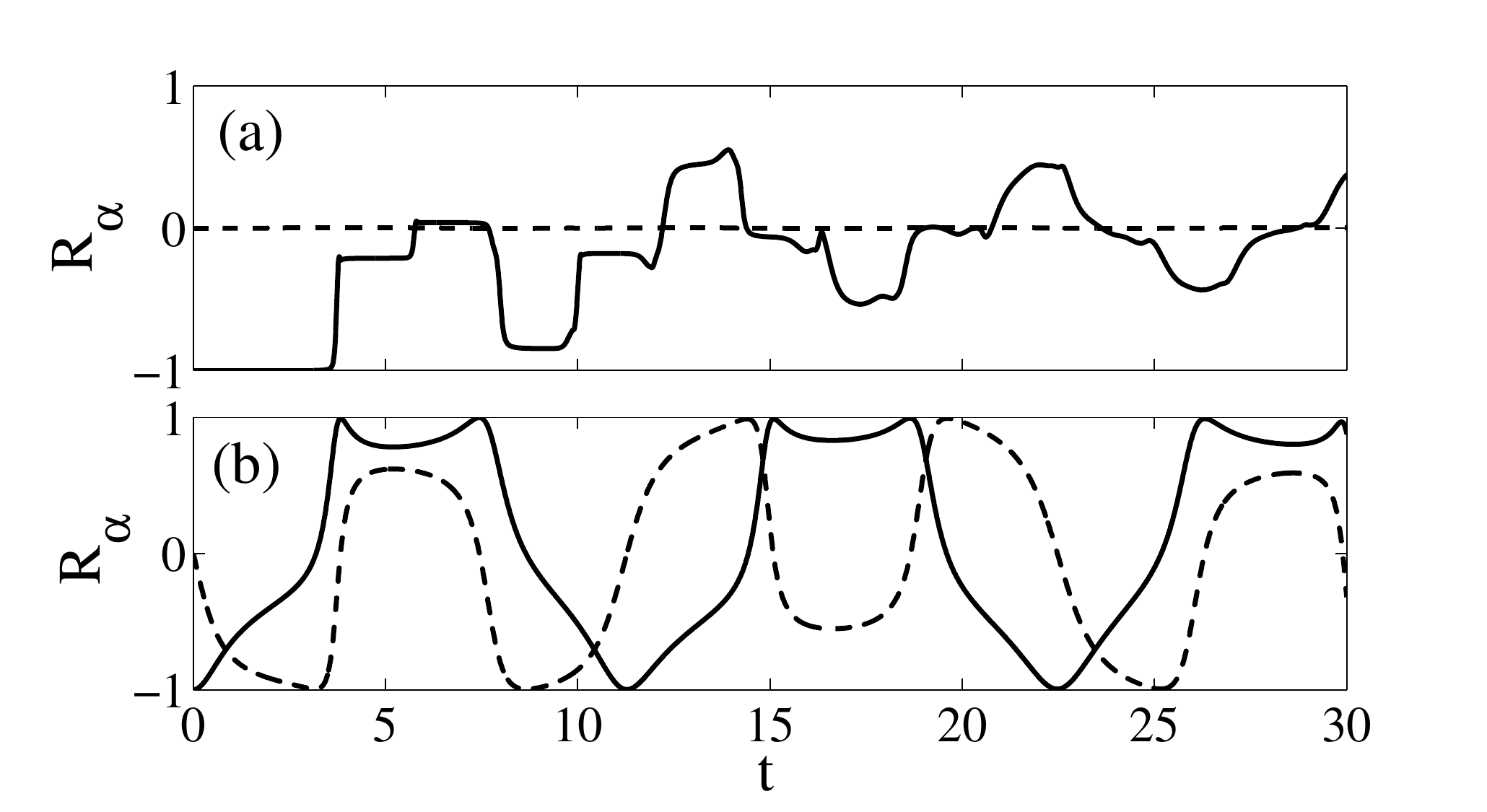}
\caption{Bloch vector components $R_x$ (dashed lines) and $R_y$ (solid lines). For the upper plot (a), the trap has been displaced in th $y$-direction, $x_s=0$ and $y_s=28$, while in the lower plot (b) the displace direction is the perpendicular, $x_s=28$ and $y_s=0$. In both figures, $v_x=10$ and $v_y=15$, and the average energy $\bar{E}\approx280$.  }
\label{fig4}
\end{figure} 

From the form of the non-adiabatic coupling~(\ref{nonadcoup}), it follows that transitions between the adiabatic states (\ref{adstate}) are restricted to the vicinity of the DP. These non-adiabatic transitions are manifested as rapid changes in the Bloch vector~(\ref{blochvec}). In Fig.~\ref{fig4} we present two examples of the Bloch vector evolution (in both examples $R_z(t)\approx0$). In Fig.~\ref{fig4} (a), the trap has been shifted in the $y$-direction. For short times, the shift of the trap induces a build-up of momentum in the opposite $y$-direction as a consequence of the Ehrenfest theorem. This adds with the non-zero $y$-component of momentum before the quench. The average momentum in the $x$-direction remains zero and as a consequence $R_x(t)\approx0$, see Eq.~(\ref{blochapp}). 

These dynamics change qualitatively if the trap is shifted in the $x$-direction instead of the $y$-direction. For sufficiently large shifts of $x_s$, the wave packet will set off along the adiabatic potentials and encircle the DP. The spin dynamics should therefore not display the same type of ``jumps'' that appear when the wave packet traverses the DP. Moreover, since the average momentum in the $x$-direction is in general non-zero, $R_x(t)$ will also be non-zero. The results are demonstrated in Fig.~\ref{fig4} (b). Compared to the first example in (a), the wave packet does not spend much time near the DP so the wave packet delocalization occurs more slowly. To a large extent the evolution is driven by harmonicity, in contrast to the example of Fig.~\ref{fig4} (a) where the anharmonicity of the Born-Huang term, and the non-adiabatic transitions near the DP, push the system away from semi-classical evolution. The figure demonstrates how the dynamics can depend on the initial conditions, in both (a) and (b), $\bar{E}\approx280$ but the wave packet broadening starts earlier in (a) than in (b). This type of state-dependence has been discussed in Ref.~\cite{fid0}; generically there is a period $t_s$ where the width of the wave packet stays nearly constant, followed by a rapid broadening. The time-scale $t_s$ depends strongly on the initial conditions, while the proceeding evolution after $t_s$ seems pretty generic for chaotic systems.   

\subsection{Long time dynamics; thermalization}\label{ssec4b}

Whenever we consider an anisotropic SO-coupling, $v_x\neq v_y$, from the Figs.~\ref{fig2} and \ref{fig3} it is clear how the adiabatic classical model becomes chaotic. Beyond the adiabatic model, it has been shown~\cite{Exbb} that the full anisotropic model, i.e., $E\times(\beta_x+\beta_y)$ Jahn-Teller model, is chaotic in the sense of {\it level repulsion}~\cite{haake} of eigenenergies. For the isotropic $E\times\varepsilon$ Jahn-Teller model, on the other hand, the level repulsion effect is not as evident, however a weak repulsion also in this model signals emergence of quantum chaos~\cite{Exechaos}.

\begin{figure}
\includegraphics[width=8cm]{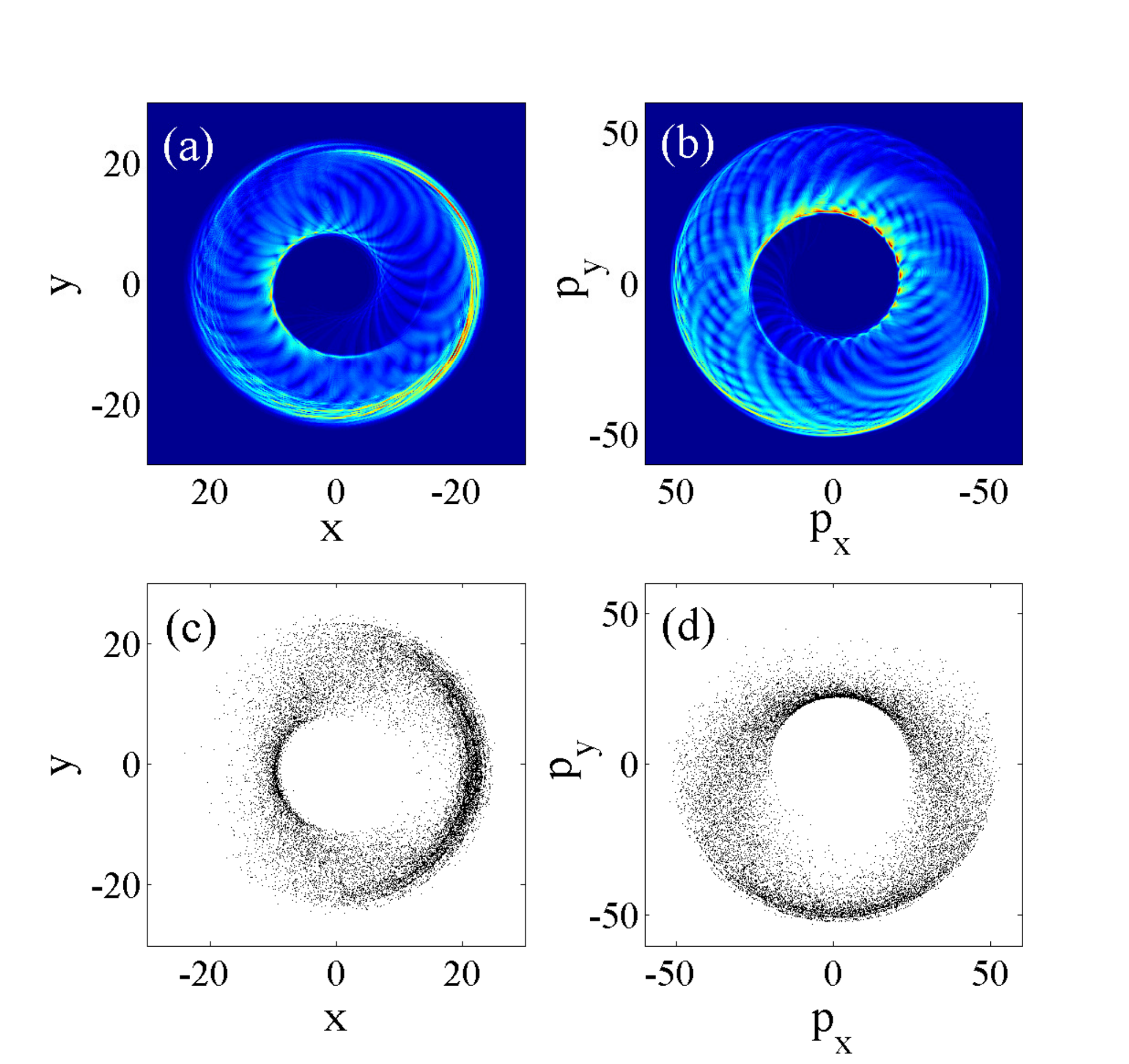}
\caption{(Color online) Distributions $|\Psi(x,y,t_f)|^2$ ((a) and (c)) and $|\Phi(p_x,p_y,t_f)|^2$ ((b) and (d)) at $t_f=400$ for the Rashba SO-coupled model. At time $t=0$, the trap is suddenly displaced from $x_0=y_0=16$ to $x_0=y_0=0$. The initial ground state is then quenched into a localized excited state. The upper two plots (a) and (b) display the results from full quantum simulations of the adiabatic model (\ref{adham}), while the lower plots (c) and (d) show the corresponding semi-classical TWA distributions. The average semi-classical energy $\bar{E}\approx-192$ with a standard deviation $\delta\bar{E}\approx22$. The dimensionless SO-coupling strengths $v_x=v_y=30$.  
}
\label{fig4}
\end{figure}

The goal of this subsection is to study the long time dynamics of the system;  specifically if equilibration occurs, and if so, does the equilibrated state mimic a thermal state. A distinguishing property of thermal states is, for example ergodicity, i.e., the distributions $|\Psi(x,y,t)|^2$ and $|\Phi(p_x,p_y,t)|^2$ spread out over their accessible energy shells. Moreover, for a thermally equilibrated state, the distributions show seemingly irregular interference structures on scales of the order of the Planck cells, which normally become even finer in the Wigner quasi distribution~\cite{zurek1,beenakker,com2}. Non-thermalized states, on the contrary, typically leave much more regular traces of quantum interference in their distributions. While such often symmetrical structures are absent for thermalized states, we will demonstrate that thermalized distributions may still show clear density fluctuations on scales larger than the Planck cells. These are examples of {\it quantum scars} and they are remnants of classical periodic orbits~\cite{scar}.  

\begin{figure}
\includegraphics[width=8cm]{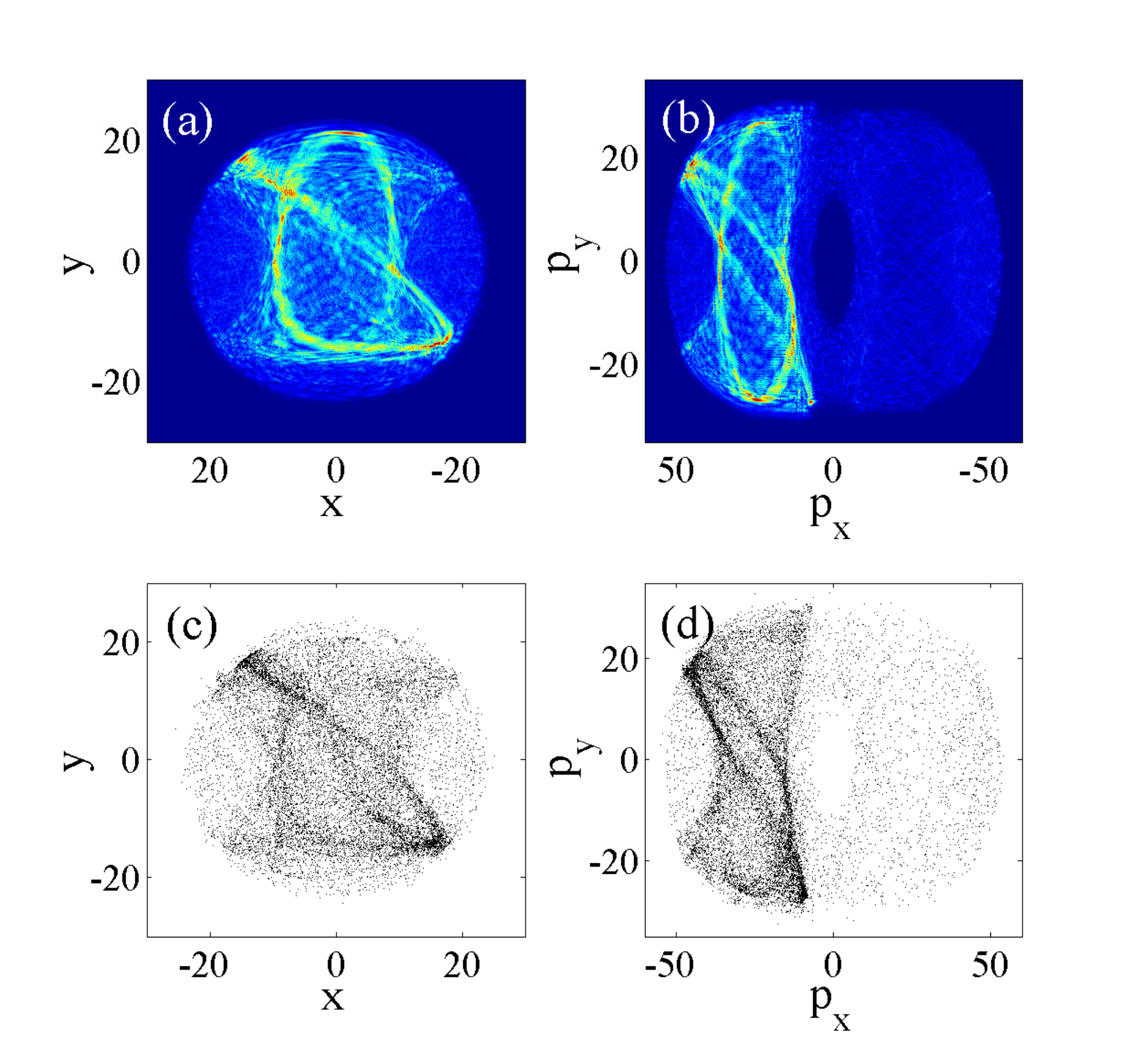}
\caption{(Color online) Same as Fig.~\ref{fig4} but for the anisotropic SO-coupled model with $v_x=20$ and $v_y=30$. The largely populated regions are so called quantum scars and derive from properties of the underlying classical model, i.e. they are not outcomes of some coherent quantum mechanism.}
\label{fig5}
\end{figure} 

We begin by considering the adiabatic isotropic model with $v_x=v_y=30$, and trap shifts $x_s=y_s=16$. After a quench of the trap position, the initial energy is $\bar{E}=\langle\hat{H}_{ad}^{(-)}\rangle\approx-192$. This energy corresponds to the energy of the Poincar\'e section presented in Fig.~\ref{fig2}. The resulting distributions are shown in Fig.~\ref{fig5} (a) and (b) after a propagation time $t_f=400$ . The final time $t_f$ approximates 60 classical oscillations. Both the real space density $|\Psi(x,y,t)|^2$ and momentum density $|\Phi(p_x,p_y,t)|^2$ reveal clear interference patterns as anticipated. The DP at the origin $(p_x,p_y)=(0,0)$ repels the wave function forming a ``hole.'' The lack of zero momentum states induces a mass flow in real space and a similar ``hole'' in its distribution. The classically energetically accessible regions are given by
\begin{equation}\label{cl_allow}
\begin{array}{l}
\displaystyle{x^2+y^2\leq 2E_{\mathrm{max}}+v_y^2},\\ \\
\displaystyle{p_x^2+p_y^2-2\sqrt{v_x^2p_x^2+v_y^2p_y^2}\leq2E_{\mathrm{max}}},
\end{array}
\end{equation}
where $E_{\mathrm{max}}$ is the maximum energy component noticeably populated by the state. 

The quantum results are compared with the TWA distributions displayed in the lower plots (c) and (d) of the same Fig.~\ref{fig5}. The same kind of ring-shape is obtained, and the concentration in density appears at the same locations for both the quantum and classical simulations. Expectedly, the quantum interference taking place within the wave packet is not captured by the TWA. This follows since single semi-classical trajectories are treated independently, i.e. added incoherently, while a quantum wave packet must be considered as one entity. For a TWA approach of the full isotropic $E\times\varepsilon$ Jahn-Teller model~(\ref{SOham}) we refer to Ref.~\cite{elham}.  

The situation is drastically changed when we break the rotational $U(1)$ symmetry by assuming $v_x\neq v_y$. The result for low initial energy is depicted in Fig.~\ref{fig5} (a) and (b). The energy is comparable to the potential barrier separating the two minima in the adiabatic potentials, and as a consequence, the wave packet is predominantly localized in the left minima. The density modulations seems now much more irregular in comparison to Fig.~\ref{fig4}. In the seemingly random density distribution, some clear density maxima emerge, both in momentum as well as in real space. These density accumulations derive from periodic orbitals of the underlying classical model and are termed quantum scars~\cite{scar,scarexp,coldscar}. The appearance of scars is an example of the classically chaotic model leaving a trace in its quantum counterpart. The scars are also captured in the semi-classical TWA, shown in Fig.~\ref{fig5} (c) and (d), supporting their classical origin.     

\begin{figure}
\includegraphics[width=8cm]{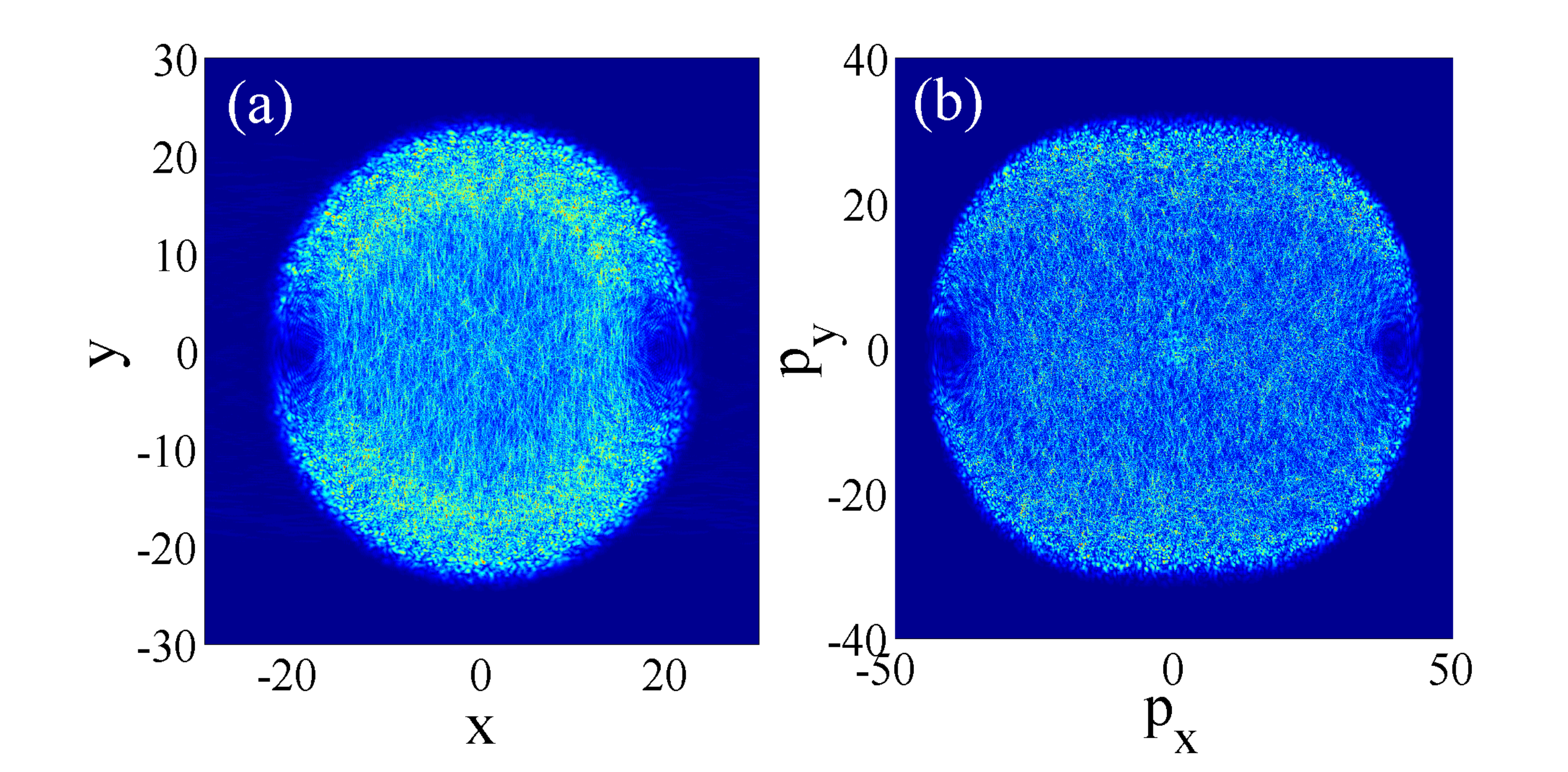}
\caption{(Color online) Same as Fig.~\ref{fig5} but for an initial energy $E>0$. The dimensionless SO-couplings $v_x=14$ and $v_y=21$, while the shifts $x_s=y_s=16$ giving an average energy $\bar{E}=\langle \hat{H}_{SO}\rangle\approx36.5$.  }
\label{fig7}
\end{figure} 

When we shift the trap for larger values on $x_s$ and $y_s$, the energy is increased and at some point the BOA breaks down. An example, obtained from integrating the full model (\ref{SOham}), is presented in Fig.~\ref{fig7}. For these higher energies there are no signs of quantum scars. As for the situation of Fig.~\ref{fig5}, the spread of the wave packet and the irregular interference patterns indicates thermalization. 

\begin{figure}
\includegraphics[width=8cm]{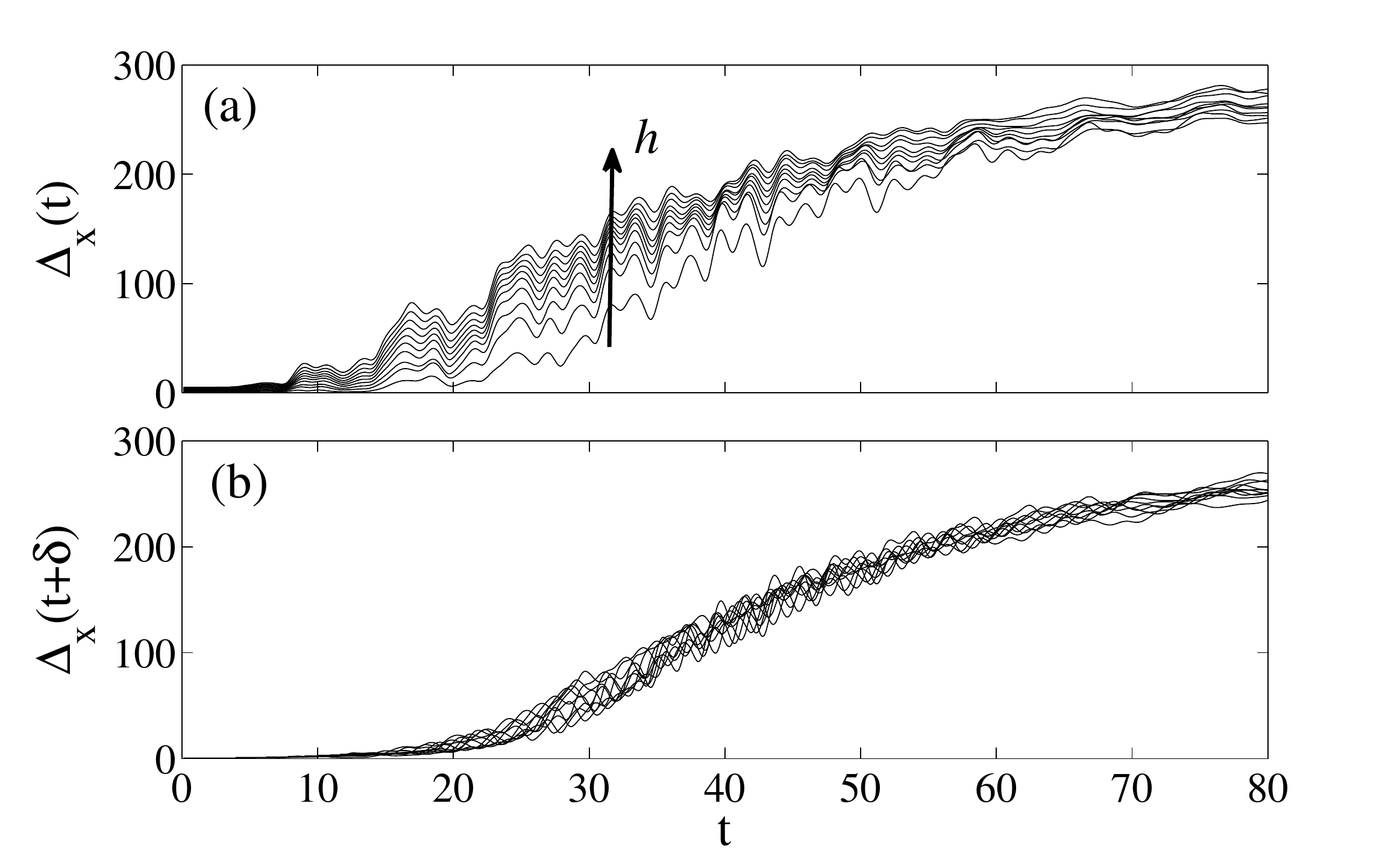}
\caption{Examples of the phase space area $\Delta_x(t)$ for different $h$-values ($h=1,\,2,\,3,\,\dots ,10$). The upper plot (a) gives $\Delta_x(t)$ without shifting the time, while for the lower one (b) time has been shifted by $\delta=\log(h)/\lambda$. The arrow indicates increasing $h$-values. It is clear how the spread in $\Delta_x(t)$ between different $h$ values is suppressed when we shift the time. The trap shifts $x_s=y_s=19$ resulting in an energy $\bar{E}\approx-88$. The maximum Lyaponov exponent $\lambda=0.18$. 
}
\label{fig9}
\end{figure}

This far we have demonstrated thermalization for the anisotropic SO coupled model, but not discussed corresponding time-scales. One related question is how the evolution of various expectation values scale with $h$ (dimensionless Planck constant). It has been argued that the Ehrenfest time, Eq.~(\ref{ehrenfest}), can be a measure of the thermalization time~\cite{altland}. We will now explore how the phase space area $\Delta_\alpha(t)=\Delta\alpha\Delta p_\alpha$ ($\alpha=x,\,y$), where $\Delta\alpha$ and $\Delta p_\alpha$ are the variances of $\hat{\alpha}$ and $\hat{p}_\alpha$ respectively, evolves for different values of $h$. Since $\Delta_x(t)$ and $\Delta_y(t)$ behave similarly we focus only on $\Delta_x(t)$. For thermalization, $\Delta_x(t)\Delta_y(t)$ is an effective measure of the covered phase space volume, and for large times $t$ it should more or less approach the accessible phase space volume as the distribution spreads over the whole energy shell. We have chosen to study $\Delta_x(t)$ since it fluctuates relatively little before reaching its asymptotic value. In Fig.~\ref{fig9} (a) we display $\Delta_x(t)$ for 10 different values on $h$ ranging from $h=1$ to $h=10$. The arrow in the plot shows the direction of increasing $h$'s. As is seen, by increasing $h$ the wave packet broadening starts earlier and the state equilibrates faster. If the Ehrenfest time $T_E$ sets the typical time scale in the process, by shifting the time with $\delta=\log(h)/\lambda$ we should recover a ``clustering'' of the curves. This is indeed verified in Fig.~\ref{fig9} (b) where the curves have been shifted in time by $\delta$. The corresponding Lyaponov exponent $\lambda$ has been optimized in order to minimize the spread in the curves. The obtained value $\lambda=0.18$ is somewhat larger than the numerically calculated one $\lambda=0.12$ but still of the same order. The picture also makes clear that the wave packet broadening kicks in after some time $t_s$ as anticipated above. 

\begin{figure}
\includegraphics[width=8cm]{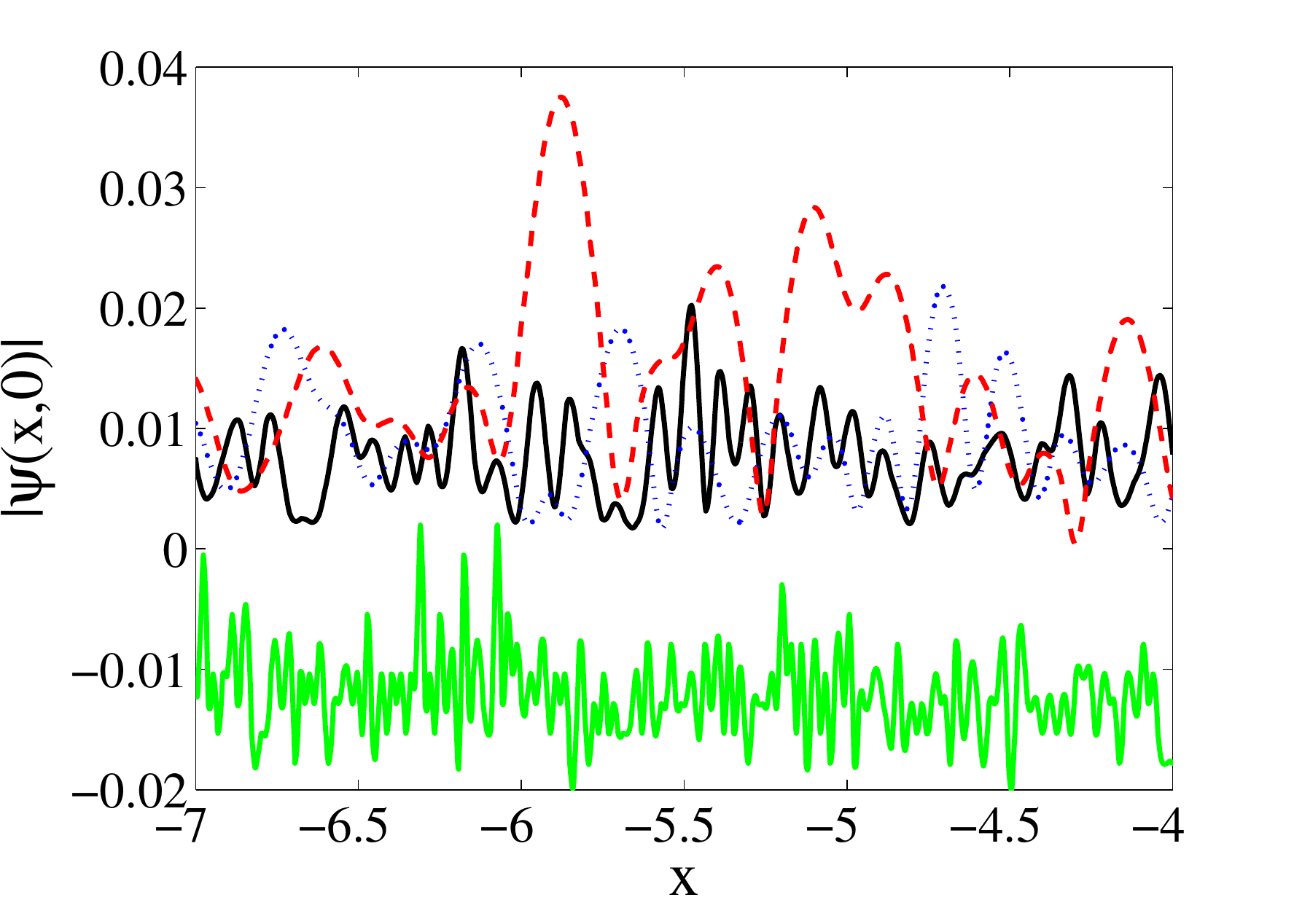}
\caption{(Color online) Sections of $|\psi(x,y=0)|$ for different values on the dimensionless Planck's constant $h$: $h=1$ (black solid line), $h=2$ (blue dotted line), and $h=3$ (red dashed line). The final time $t_f=80$, $x_s=y_s=16$, and $v_x=14$ and $v_y=20$. As a comparison between classical and quantum results, we also include the TWA results as a green solid line, calculated for $h=1$. The green line has been shifted downward with 0.02 for clarity.  
}
\label{fig9b}
\end{figure}

The route to thermalization can typically be divided into; $(i)$ a classical drift, and $(ii)$ quantum diffusion~\cite{altland}. The role of the quantum diffusion for thermalization was analyzed in Ref.~\cite{altland}, where it was found to ``smoothen'' the phase space distributions preventing sub-Planck structures. For the classical drift there is no lower bound on the fineness of density structures that can form, and characteristic for classical chaotic dynamics is that ever finer formations build-up as a result of the typical ``stretching-and-folding'' mechanism. However, in a quantum chaotic system, when the structures reach the Planck cell regime, the quantum pressure becomes too strong and the quantum diffusion then prevents any further structures to form. Thus, Planck's constant sets a lower bound for the fluctuations in the distributions. This quantum smoothening is demonstrated in Fig.~\ref{fig9b}, where we plot a section of $|\psi(x,y=0)|$ for different values on the scaled dimensionless Planck's constant $h$ ($=1,\,2,\,3$ for black, blue, and red lines respectively). The effect is clearly seen in the figure. A similar pattern is found (not shown) also for the momentum distributions. For the classical system, corresponding to $h=0$, there is no lower limit on how fine the structures can be. We indicate this by also plotting the TWA results in the same figure as a green line (note that the green line has been shifted downward in order to separate it from the quantum results). The number of trajectories used for the figure is 250 000, and if we would like to produce finer structures (by propagating the system for longer times) we would need many more trajectories and the simulation would rapidly become very time consuming. 

Related to the above discussion a note on quantum phase space distributions is in order. It is well known that sub-Planck structures are common in the Wigner distribution~\cite{zurek1}. This is not contradicting any quantum uncertainty relation. After all, the Wigner distribution is not a proper probability distribution, despite the fact that its marginal distributions reproduce the correct real and momentum space probability distributions. The Husimi $Q$-function, while not possessing the proper marginal distributions, is positive definite and lacking singularities, and it is indeed found that the $Q$-function does not support sub-Planck structures~\cite{duncan}.   

\begin{figure}
\includegraphics[width=8cm]{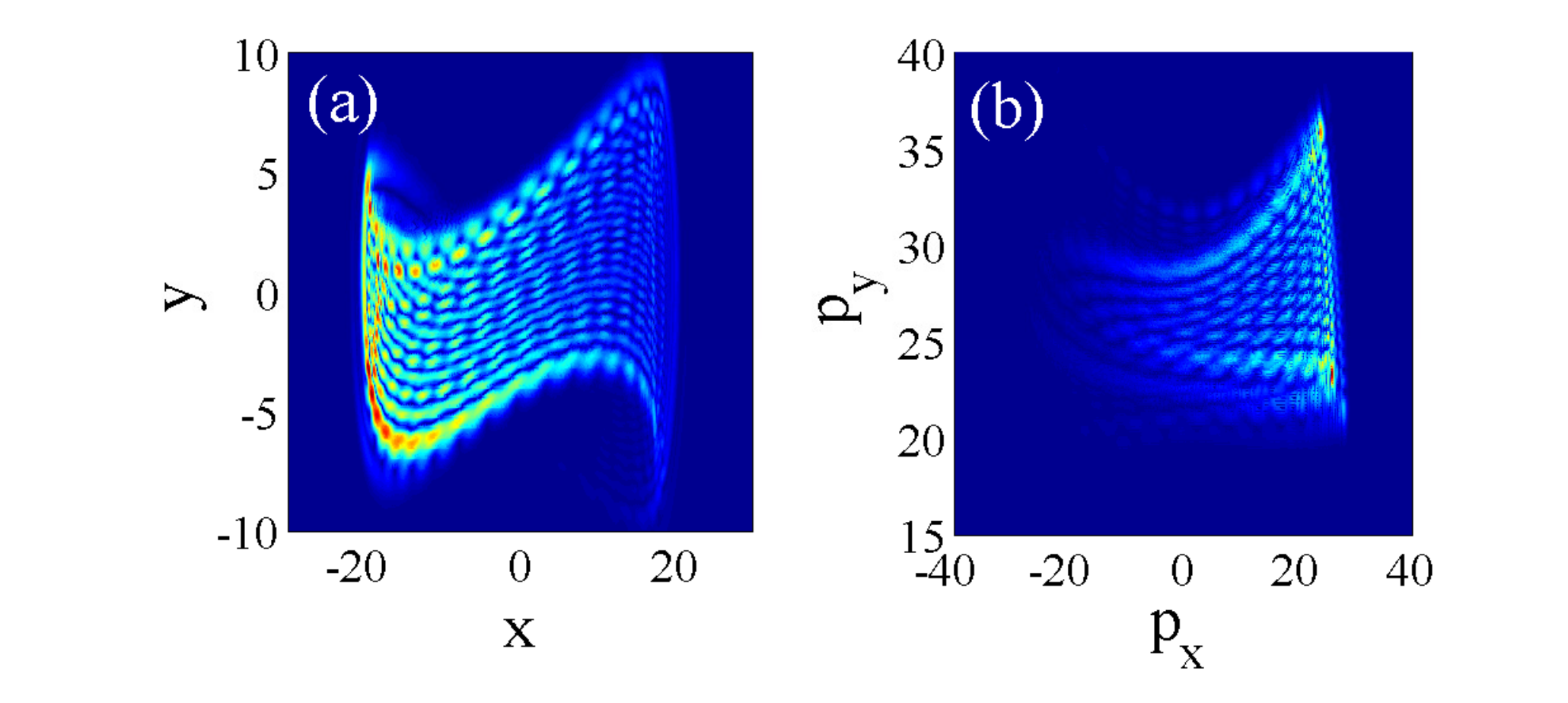}
\caption{(Color online) Same as Fig.~\ref{fig5} but for the shifts $x_s=20$ and $y_s=0$. 
For the given dimensionless parameters, the initial state is such that its dynamics should be regular according to the corresponding Poincar\'e section, Fig.~\ref{fig3}. The energy $\bar{E}\approx-250$.}
\label{fig10}
\end{figure}

We finish this subsection by analyzing the dynamics in the islands of the Poincar\'e sections of Fig.~\ref{fig3} where the classical theory predicts regular evolution. From Fig.~\ref{fig3} (c) we have that for $p_x\approx20$ and $x\approx y\approx0$ the evolution should be regular. We can achieve such a situation by using the quench-shifts $x_s=20$ and $y_s=0$. As for the examples above, we propagate the state for a time $t_f=400$, and the resulting distributions are given in Fig.~\ref{fig10}. The striking difference with Figs.~\ref{fig5} and \ref{fig7} is evident; no irregular structure is apparent, but clear regular interference patterns are. We have verified that the interference structure prevails also after doubling the time, $t_f=800$. 

\subsection{Proposed experimental realization}\label{ssec4c}

Much of the above dynamics can be observed in a system of cold atoms with synthetic SO-coupling, for example, a system of ${}^{87}\textrm{Rb}$ with a synthetic field induced by the 4-level scheme~\cite{Nlevel}. In this system, the recoil energy $E_r = mv^2 \sim \hbar \times 50 \, \textrm{kHz}$. The synthetic field limits the lifetime of the experiment to $t_l \sim 1 s$~\cite{spielman1,spielman2}. To push the experiment into the long time regime, we will use a trapping frequency of $\omega / 2\pi = 30 \,  \textrm{Hz}$. These parameters will give a dimensionless value of $v_y = \sqrt{\frac{E_r}{\hbar\omega}} \sim 11$, with $v_x$ tunable between $0$ and $11$. The large trapping frequency will provide a sufficient number of oscillations for thermalization to occur. We could consider values of $v_y \sim 30$ by decreasing the trapping frequency to $10 \textrm{Hz}$, but then the lifetime of the system may be at the boarder for thermalization.

The condensate can be adiabatically loaded to one of the two states at the bottom of the momentum-space potential, defined by ${\bf p} = \pm m v_y \hat{y}$. The quench can then be preformed by shifting the minimum of the real-space trapping potential. We then let the system evolve until we reach either the thermalization time, or the lifetime of the experiment. The momentum distribution can be measured with a destructive time-of-flight (TOF) measurement~\cite{spielman1,spielman2}, which should reveal thermalization as well as signatures of quantum scars. Repeated experimental measurements allow for time-resolved calculation of expectation values. Similarly, the quantum spin jumps near the DP, as discussed in Sec.~\ref{ssec4a}, can be observed using a spin-resolved  TOF measurement.

As a final remark, for a weakly interacting gas we work near a Feschbach resonance~\cite{Feschbach}. However, for realistic parameters~\cite{spiel}, we estimate a scattering length $a_s\sim3\times10^{-9}$ m, $N\sim5\times10^5$ atoms, and a transverse harmonic trapping frequency $\omega_z\sim100$ Hz. For these parameters, the characteristic scale of the non-linearity is $\mu \sim h\times 1 kHz$, which is smaller than the recoil energy above, suggesting  the non-linear term will play only a minor role. We have numerically verified that the results do not change qualitatively by solving the corresponding non-linear Gross-Pitaevskii equation. Indeed, we find the deviations with a non-linearity are not large enough to be seen by eye.

\section{Conclusions}\label{sec5}
In this paper we studied dynamics, deriving from a quantum quench, in anisotropic SO-coupled cold gases, focusing primary on aspects arising from the fact that the underlying classical model is chaotic. The evolution of the initially localized wave packet on its way to equilibration has been analyzed, and we have shown how a classical period of limited spreading is followed by a collapse regime dominated by rapid spreading. After the collapse period, the wave packet is maximally delocalized, but still possesses quantum interference structures. At the Ehrenfest time, the state has approximately equilibrated as is seen in the decay of expectation values, as well as seemingly irregular density fluctuations both in real and momentum space. We showed that the fine structure of these fluctuations are limited by the quantum diffusion, and thereby the size of the Planck's constant $h$. For the isotropic model, after the collapse no thermalization is found, as is expected from the integrability of the underlying classical model. 

For smaller energies, when the wave packet predominantly populates one of the dual potential wells, thermalization is again seen. Here, however, an additional phenomenon appears in terms of quantum scars. These density enhancements emerge along classically periodic orbits. They are classical in nature and long lived. Quantum scars have also been studied in different cold atom settings; atoms in an optical lattice and confined in an anisotropic harmonic trap~\cite{coldscar}. The results on thermalization presented in this work is most likely also applicable to the set-up of Ref.~\cite{coldscar}. We also demonstrated that for certain fine tuned initial states, the dynamics stays regular even in the anisotropic model. In the classical picture, these solutions correspond to the ones belonging to regular islands in the otherwise chaotic Poincar\'e sections.  

We argue that the present system is ideal for studies of quantum chaos and quantum thermalization for numerous reasons. The system parameters can be tuned externally by adjusting the wavelength of the lasers inducing the SO-coupling, and as we discussed in Sec.~\ref{ssec4c} the SO dominated regime is reachable in current experiments. Moreover, both state preparation and detection are relatively easily performed in these setups. Equally important, the system is well isolated from any environment and coherent dynamics can be established up to hundreds of oscillations which is well beyond the themralization time. The energy of the state is simply controlled by the trap displacement, and it should for example be possible to give the system small energies such that the atoms reside mainly in one potential well where quantum scars develop. 

We finish by pointing out that the present model is also different from most earlier studies on quantum thermalization~\cite{polkovnikov,thermo2} in the sense that the dynamics is essentially ``single-particle'' and not arising from many-body physics. Related to this, we have numerically verified that adding a non-linear term $g|\Psi(x,y,t)|^2$ to the Hamiltonian does not change our results qualitatively for moderate realistic interaction strengths $g$. In order to enter into the regime where interaction starts to affect the results, one would need a condensate with a large number of atoms ($\sim$millions of atoms) or alternatively externally tune the scattering length via the method of Feshback resonances.   

\begin{acknowledgments}
The authors thank Ian Spielman for helpful comments. SFB/TR 12 is acknowledged for financial support. JL acknowledges Vetenskapsr\aa det (VR), DAAD (Deutscher Akademischer Austausch Dienst), and the Royal Research Council Sweden (KVA) for financial help. BA acknowledges the sponsorship of the US Department of Commerce, National Institute of Standards and Technology, and was supported by the National Science Foundation under Physics Frontiers Center Grant PHY-0822671 and by the ARO under the DARPA OLE program.
\end{acknowledgments}


\begin{thebibliography}{999}
\bibitem{com2} Normally, the sub-Planck cell structures are studied in the phase space distributions. However, there exist several different phase psace distributions having different properties. For example, the sub-Planck cell structures typically appear in the Wigner distributions, while in the Husimi $Q$-function, being strictly positive, small fluctuations have been ``averaged out'' and it is normally much smoother than the Wigner distribution.
\end{thebibliography}
\end{document}